\documentclass[prc,twocolumn,showpacs,amsmath,amssymb]{revtex4}
\usepackage{times}
\usepackage{graphicx}
\usepackage{dcolumn}
\usepackage{bm}
\usepackage{amstext}

\begin{document}

\title{\Large Large deformed structures in $Ne-S$ nuclei near neutron 
drip-line}
\author{S.K. Patra and C.R. Praharaj}
\affiliation{Institute of Physics, Sachivalaya Marg, Bhubaneswar-751 005, 
India} 

\date{\today}
\begin{abstract}

The structure of Ne, Na, Mg, Al, Si and S nuclei near the neutron drip-line
region is investigated in the frame-work of relativistic meannfield (RMF) 
and non-relativistic Skyrme Hartree-Fock formalisms. The drip-line of these
nuclei are pointed out. We analysed the large deformation structures and
 many of these neutron rich nuclei are quite deformed.
New magic number are seen for these deformed nuclei.
\end{abstract}
\maketitle

\section{Introduction}

The structure of light nuclei near the neutron drip-line is very interesting
for a good number of exotic phenomena. Nuclei in this region
are very different in collectivity and clustering features than the stable
counterpart in the nuclear chart.  For example, the neutron magicity
is lost for the N=8 nucleus for $^{12}$Be \cite{navin00} and N=20 for $^{32}$Mg
\cite{motobayashi95}.
The discovery of large collectivity of $^{34}$Mg by Iwasaki et al. 
\cite{iwasaki01} is
another example of such properties. The deformed structures and core
excitations of Mg and neighboring nuclei and location of drip-line in
this mass region is an important matter \cite{patra91}.
On the other hand,
the appearance of N=16 magic number for $^{24}$O is well established
\cite{ozawa00}. The
discovery of the two isotopes $^{40}$Mg and $^{42}$Al, once predicted
to be drip-line nuclei \cite{baumann07,moller95} gives indication that the 
neutron drip-line is
located towards the heavier mass region.
The existence of neutron halo in
$^{11}$Li is well established and the possibility of proton halo in $^{8}$B
and the neutron halo in $^{14}$Be and $^{17}$B are very interesting
phenomena for the drip-line nuclei. In addition to these above exciting
properties, the cluster structure of entire light mass nuclei and the
skin formation in neutron-drip nuclei provide us features for the study of light mass
drip-line nuclei. Also, the exotic neutron drip-line nuclei play a role in 
many astrophysical studies.
In this paper, our aim is
to study the neutron drip-line for Ne-S isotopic chain in the frame-work
of a relativistic mean field (RMF) and nonrelativistic Skyrme Hartree-Fock
formalism and analyse the large deformation of 
these isotopes.

The paper is organised as follows: The relativistic and 
non-relativistic mean field formalisms are described very briefly
in Section II. The results obtained from the relativistic mean field 
(RMF) and Skyrme-Hartree-Fock (SHF) formalisms, and a discussion of 
these results, are presented in Section III. Finally summary 
and concluding remarks are given in Section IV.

\section{Theoretical framework}

Mean field methods have been widely used in the study of binding 
energies and other properties of nuclei \cite{vautherin72,reinhard89}. 
Although the older version of the SHF and RMF models have some
limitation to reproduce some of the observables, the recent formalisms
are quite efficient to predict the
bulk properties of nuclei not only near the stability valley, but
also for the nuclei near the proton and neutron drip-lines.
We use here two of the successful mean field models \cite{vautherin72,
rei95,cha97,cha98,stone07,stone03,sero86,ring90}
(Skyrme Hartree-Fock and the Relativistic Mean
Field ) to learn about the properties of drip-line nuclei $Ne-S$. 

\subsection{The Skyrme Hartree-Fock (SHF) Method}

There are many known parametrizations of Skyrme interaction which reproduce the 
experimental data for ground-state properties of finite nuclei and for the 
observables of infinite nuclear matter at saturation densities, giving more or 
less comparable agreements with the experimental or expected empirical
data. The general form of the Skyrme effective interaction, used in the 
mean-field models, can be expressed as a density functional $\cal H$ 
\cite{cha97,stone07}, given as a function of some empirical parameters, as
\begin{equation}
{\mathcal H}={\mathcal K}+{\mathcal H}_0+{\mathcal H}_3+ {\mathcal H}_{eff}+\cdots \label{eq:1}
\end{equation}
where ${\cal K}$ is the kinetic energy term, ${\cal H}_0$ the zero range, ${\cal H}_3$ the 
density dependent and ${\cal H}_{eff}$ the effective-mass dependent terms, which are 
relevant for calculating the properties of nuclear matter. These are functions of 9 parameters 
$t_i$, $x_i$ ($i=0,1,2,3$) and $\eta$, and are given as
\begin{eqnarray}
{\mathcal H}_0&=&\frac{1}{4}t_0\left[(2+x_0)\rho^2 - (2x_0+1)(\rho_p^2+\rho_n^2)\right],
\label{eq:2}
\\
{\mathcal H}_3&=&\frac{1}{24}t_3\rho^\eta \left[(2+x_3)\rho^2 - (2x_3+1)(\rho_p^2+\rho_n^2)\right],
\label{eq:3}
\\
{\mathcal H}_{eff}&=&\frac{1}{8}\left[t_1(2+x_1)+t_2(2+x_2)\right]\tau \rho \nonumber \\
&&+\frac{1}{8}\left[t_2(2x_2+1)-t_1(2x_1+1)\right](\tau_p \rho_p+\tau_n \rho_n). \nonumber \\
\label{eq:4}
\end{eqnarray}
The kinetic energy ${\cal K}=\frac{\hbar^2}{2m}\tau$, a form used in the Fermi gas model for 
non-interacting fermions. The other terms, representing the surface contributions of a 
finite nucleus with $b_4$ and $b^{\prime}_4$ as additional parameters, are
\begin{eqnarray}
{\mathcal H}_{S\rho}&=&\frac{1}{16}\left[3t_1(1+\frac{1}{2}x_1)-t_2(1+\frac{1}{2}x_2)\right](\vec{\nabla}\rho)^2 \nonumber\\
&&-\frac{1}{16}\left[3t_1(x_1+\frac{1}{2})+t_2(x_2+\frac{1}{2})\right] \nonumber\\
&&\times\left[(\vec{\nabla}\rho_n)^2+(\vec{\nabla}\rho_p)^2\right],
\text{ and}
\label{eq:5}
\\
{\mathcal H}_{S\vec{J}}&=&-\frac{1}{2}\left[{b_4}\rho\vec{\nabla}\cdot\vec{J}+{b^{\prime}_4}(\rho_n\vec{\nabla}\cdot\vec{J_n}
+\rho_p\vec{\nabla}\cdot\vec{J_p})\right].
\label{eq:6}
\end{eqnarray}
Here, the total nucleon number density $\rho=\rho_n+\rho_p$, the kinetic energy density 
$\tau=\tau_n+\tau_p$, and the spin-orbit density $\vec{J}=\vec{J}_n+\vec{J}_p$. The 
subscripts $n$ and $p$ refer to neutron and proton, respectively, and $m$ is the nucleon 
mass. The $\vec{J}_q=0$, $q=n$ or $p$, for spin-saturated nuclei, i.e., for nuclei 
with major oscillator shells completely filled. The total binding energy (BE) of a 
nucleus is the integral of the density functional $\cal H$.

At least eighty-seven parametrizations of the Skyrme interaction are published since 1972
\cite{stone03} where $b_4=b^{\prime}_4=W_0$,  we have used here the Skyrme SkI4 set with $b_4\ne b^{\prime}_4$
\cite{rei95}. This parameter set is designed for considerations of proper spin-orbit interaction in finite nuclei,
related to the isotope shifts in Pb region and is better suited for the
study of exotic nuclei. Several more recent Skyrme parameters such as 
SLy1-10, SkX, SkI5 and SkI6 are obtained by fitting the Hartree-Fock (HF) results with experimental 
data for nuclei starting from the valley of stability to neutron and
proton drip-lines \cite{cha97,rei95,cha98,brown98}. 

\subsection{The Relativistic Mean Field (RMF) Method}
The relativistic mean field approach is well-known and the theory is well documented
\cite{sero86,ring90}. Here we start with the relativistic Lagrangian
density for a nucleon-meson many-body system, as
\begin{eqnarray}
{\cal L}&=&\overline{\psi_{i}}\{i\gamma^{\mu}
\partial_{\mu}-M\}\psi_{i}
+{\frac12}\partial^{\mu}\sigma\partial_{\mu}\sigma
-{\frac12}m_{\sigma}^{2}\sigma^{2}\nonumber\\
&& -{\frac13}g_{2}\sigma^{3} -{\frac14}g_{3}\sigma^{4}
-g_{s}\overline{\psi_{i}}\psi_{i}\sigma-{\frac14}\Omega^{\mu\nu}
\Omega_{\mu\nu}\nonumber\\
&&+{\frac12}m_{w}^{2}V^{\mu}V_{\mu}
+{\frac14}c_{3}(V_{\mu}V^{\mu})^{2} -g_{w}\overline\psi_{i}
\gamma^{\mu}\psi_{i}
V_{\mu}\nonumber\\
&&-{\frac14}\vec{B}^{\mu\nu}.\vec{B}_{\mu\nu}+{\frac12}m_{\rho}^{2}{\vec
R^{\mu}} .{\vec{R}_{\mu}}
-g_{\rho}\overline\psi_{i}\gamma^{\mu}\vec{\tau}\psi_{i}.\vec
{R^{\mu}}\nonumber\\
&&-{\frac14}F^{\mu\nu}F_{\mu\nu}-e\overline\psi_{i}
\gamma^{\mu}\frac{\left(1-\tau_{3i}\right)}{2}\psi_{i}A_{\mu} .
\end{eqnarray}
All the quantities have their usual well known meanings. From the relativistic 
Lagrangian we obtain the field equations for the nucleons and mesons. These 
equations are solved by expanding the upper and lower components of the Dirac 
spinors and the boson fields in an axially deformed harmonic oscillator basis 
with an initial deformation. The set of coupled equations is solved numerically 
by a self-consistent iteration method. 

The centre-of-mass motion energy correction is estimated by the usual harmonic 
oscillator formula $E_{c.m.}=\frac{3}{4}(41A^{-1/3})$. The constant gap
BCS pairing is used to add the pairing effects for the open shell nuclei.
It is to be noted that in the present work only intrinsic state solutions
are presented. Each of these deformed intrinsic states is a superposition
of various angular momenta states. To obtain the good angular momentum states 
and spectroscopic predictions for these nuclei near neutron drip-line 
we need to project out states of good angular momenta.
Such calculation will be considered as a future extension of this work.
The quadrupole moment deformation parameter $\beta_2$ is evaluated from the resulting proton
and neutron quadrupole moments, as $Q=Q_n+Q_p=\sqrt{\frac{16\pi}5} (\frac3{4\pi} AR^2\beta_2)$. 
The root mean square (rms) matter radius is defined as $\langle r_m^2\rangle={1\over{A}}\int\rho(r_{\perp},z) r^2d\tau$; 
here $A$ is the mass number, and $\rho(r_{\perp},z)$ is the deformed density. 
The total binding energy and other observables are also
obtained by using the standard relations, given in \cite{ring90}. We use the well 
known NL3 parameter set \cite{lala97}. This set not only reproduces the properties 
of stable nuclei but also well predicts for those far from the
$\beta$-stability valley. Also, the isoscalar monopole energy agrees excellently with 
the experimental values for different regions of the Periodic Table. The measured 
superdeformed minimum in $^{194}$Hg is 6.02 MeV above the ground state, 
whereas in RMF calculation with NL3 set, this number is 5.99 MeV \cite{lala97}. All
these facts give us confidence to use this older, though very much still in use, NL3 
set for the present investigation.

\section{Results and Discussion}

\subsection{Ground state properties from the SHF and RMF models}

There exists a number of parameter sets for the standard SHF and RMF
Hamiltonians and  Lagrangians. In some of our previous papers and of 
other authors
\cite{ring90,lala97,patra1,patra2,patra3}
the ground state properties, like the binding energies (BE),
quadrupole moment deformation parameters $\beta_2$, charge radii ($r_c$) 
and other bulk properties are evaluated
by using the various non-relativistic and relativistic parameter sets. 
It is found that, more or less, most of the recent parameters reproduce well the
ground state properties, not only of stable normal nuclei but also
of exotic nuclei which are away from the valley of beta-stability.
So, if one uses a reasonably acceptable parameter set the prediction
of the results will remain nearly force independent. This is valid both
for SHF and RMF formalisms. However, with a 
careful inspection of these parametrizations, some of the SHF and RMF
sets can not reproduce the  empirical data. In this
context we can cite the deviation of isotopic shifts than the
experimental data \cite{sharma93} for Pb nuclei while
using SHF forces like, SkM* values \cite{tajima93}. However, the
RMF sets reproduce the kink quite nicely \cite{sharma93a}.  On the
other hand, most of the RMF sets over estimate the nuclear matter
incompressibility. In general, the predictive power of both the 
formalisms are reasonably well and can be comparable to each other,
which can be seen in the subsequent subsections. In addition to this, 
the general results SHF (SkI4) and RMF (NL3) forces are similar for 
the considered region. Thus in the subsequent results during our 
discussion we will refer the results of RMF (NL3) calculations, 
except some specific cases. Thus the result of SHF (SkI4) are 
not displayed in Tables. 

\subsection{Binding energy and neutron drip-line}
The ground state binding energy (BE) are calculated for
Ne, Na, Mg, Al, Si and S isotopes near the neutron drip-line.
This is done by comparing the prolate, oblate and spherical solution
of binding energy for a particular nucleus. For a given nucleus, the 
maximum binding energy corresponds to the ground state and other 
solutions are obtained as various excited intrinsic states. 
In Table I, the ground state binding energy for the heaviest
isotopes for the nuclei discussed are compared with the experimental
data \cite{audi03}. From the Table it is observed that the calculated binding
energies are comparable with SHF and RMF results. 
\begin{table}
\caption{\label{tab:table1}{\it The calculated ground state binding energy
obtained from SHF and RMF theory are compared with the experimentally
known heaviest isotope for Ne, Na, Mg, Al, Si and S \cite{audi03}.}
}
\begin{tabular}{|c|c|c|c|c|c|c|c|c|}
\hline
nucleus &RMF&SHF&Expt.&nucleus&RMF&SHF&Expt.\\
\hline
$^{30}$Ne& 215.1 & 210.6 & 211.2& $^{33}$Na & 237.9&234.5 &232.8 \\
$^{34}$Mg&257.7& 255.1 & 256.2&$^{39}$Al &285.2 &281.9&283.1 \\
$^{41}$Si&310.1&305.1&306.6 & $^{45}$S &353.5 &350.2&354.2  \\
\hline
\end{tabular}
\end{table}
We have listed the neutron drip-lines in Table II, which are obtained from the 
ground state binding energy for neutron rich Ne, Na, Mg, Al, Si and S nuclei.
The nuclei with the largest neutron numbers so far  experimentally
detected in an isotopic chain till date, known as experimental
neutron drip-line  are also displayed in this  Table for comparison.
The numbers given in the parenthesis are the experimentally extrapolated
values\cite{audi03}. To get a qualitative understanding of the prediction
of neutron drip-line, we have compared our results with the infinite
nuclear matter (INM) \cite{sn} and finite range droplet model (FRDM)
\cite{frdm} mass estimation. From the table, it is clear that all the predictions
for neutron drip-line are comparable to each other.
\begin{table}
\caption{\label{tab:table1}{\it The predicted neutron drip-line for Ne,
Na, Mg, Al, Si and S in RMF (NL3)  and SHF (SKI4) parameter sets are 
compared with prediction of infinite nuclear matter (INM) mass model \cite{sn}, 
finite range droplet model (FRDM) \cite{frdm} and experimental
data \cite{audi03} along with the number shown in parenthesis are the 
experimentally extrapolated values.  }}
\begin{tabular}{|c|c|c|c|c|c|c|c|c|}
\hline
nucleus &RMF&SHF&INM&FRDM&Expt.\\
\hline
Ne& 34 & 34 &36&32&30 (34) \\
Na &40 &37 &40&36&33 (37)  \\
Mg& 40 & 40 &46&40&34 (40) \\
Al &48 &48 &48&45&39 (42)  \\
Si& 54 & 48 &50&48&41 (44) \\
S  &55 &55 &53&50&45 (49)  \\
\hline
\end{tabular}
\end{table}

\begin{table}
\caption{\label{tab:table3}{\it The calculated value of charge radius ($r_{ch}$),
quadrupole moment deformation parameter $\beta_2$ and binding energy (BE) for
Ne, Na and Mg  nuclei in RMF (NL3) formalism. The maximum binding energy is the
ground state solution and all other values are the intrinsic excited state
solution. The radius $r_{ch}$ is in fm and the binding energy is in MeV.
}}
\begin{tabular}{|c|c|c|c|c|c|c|c|c|c|}
\hline
\hline
nucleus &$r_{ch}$&$\beta_2$&BE(MeV)&nucleus &$r_{ch}$&$\beta_2$&BE(MeV)\\
\hline

$^{20}$Ne & 2.970 & 0.535 & 156.7 & $^{20}$Ne & 2.901 &-0.244  & 152.0 \\
$^{21}$Ne & 2.953 & 0.516 & 165.9 & $^{21}$Ne & 2.889 &-0.241  & 161.1 \\
$^{22}$Ne & 2.940 & 0.502 & 175.7 & $^{22}$Ne & 2.881 &-0.242  & 170.5 \\
$^{23}$Ne & 2.913 & 0.386 & 181.8 & $^{23}$Ne & 2.880 &-0.249  & 179.7\\
$^{24}$Ne & 2.890 & 0.278 & 188.9 & $^{24}$Ne & 2.881 & -0.259 & 188.9\\
$^{25}$Ne & 2.907 & 0.272 & 194.2 & $^{25}$Ne & 2.886 & -0.206 & 194.2\\
$^{26}$Ne & 2.926 & 0.277 & 199.9 & $^{26}$Ne & 2.893 & -0.159 & 199.9\\
$^{27}$Ne & 2.945 & 0.247 & 203.9 & $^{27}$Ne & 2.925 & -0.183 & 203.9\\
$^{28}$Ne & 2.965 & 0.225 & 208.2 & $^{28}$Ne & 2.957 & -0.203 & 208.2\\
$^{29}$Ne & 2.981 & 0.161 & 211.2 & $^{29}$Ne & 2.974 & -0.133 & 211.2\\
$^{30}$Ne & 2.998 & 0.100 & 215.0 & $^{30}$Ne & 2.995 & -0.081 & 215.0\\
$^{31}$Ne & 3.031 & 0.244 & 216.0 & $^{31}$Ne & 3.013 & -0.133 & 216.0\\
$^{32}$Ne & 3.071 & 0.373 & 218.6 & $^{32}$Ne & 3.033 & -0.180 & 218.6\\
$^{33}$Ne & 3.095 & 0.424 & 219.5 & $^{33}$Ne & 3.044 & -0.230 & 219.5\\
$^{34}$Ne & 3.119 & 0.473 & 220.9 & $^{34}$Ne & 3.054 & -0.275 & 220.9\\
$^{35}$Ne & 3.132 & 0.505 & 220.4 & $^{35}$Ne & 3.064 & -0.315 & 215.7\\ 
$^{36}$Ne & 3.146 & 0.539 & 220.3 & $^{36}$Ne & 3.075 & -0.352 & 220.4\\
$^{24}$Na & 2.939 &- 0.250 & 189.4 & $^{24}$Na & 2.964 & 0.379 & 192.3\\
$^{25}$Na & 2.938 & -0.258 & 200.3 & $^{25}$Na & 2.937 & 0.273 & 200.6\\
$^{26}$Na & 2.940 & -0.202 & 206.3 & $^{26}$Na & 2.965 & 0.295 & 207.1\\
$^{27}$Na & 2.946 & -0.157 & 212.5 & $^{27}$Na & 2.993 & 0.323 & 214.2\\
$^{28}$Na & 2.980 & -0.184 & 217.7 & $^{28}$Na & 2.993 & 0.272 & 219.0\\
$^{29}$Na & 3.012 & -0.205 & 223.4 & $^{29}$Na & 3.004 & 0.232 & 224.3\\
$^{30}$Na & 3.025 & -0.131 & 227.5 & $^{30}$Na & 3.031 & 0.169 & 228.1\\
$^{31}$Na & 3.043 & -0.074 & 232.5 & $^{31}$Na & 3.047 & 0.108 & 232.7\\
$^{32}$Na & 3.061 & -0.129 & 233.0 & $^{32}$Na & 3.077 & 0.237 & 234.5\\
$^{33}$Na & 3.082 & -0.179 & 234.3 & $^{33}$Na & 3.113 & 0.356 & 237.9\\
$^{34}$Na & 3.095 & -0.226 & 234.8 & $^{34}$Na & 3.137 & 0.404 & 239.8\\
$^{35}$Na & 3.108 & -0.270 & 235.9 & $^{35}$Na & 3.161 & 0.450 & 242.3\\
$^{36}$Na & 3.121 & -0.308 & 236.9 & $^{36}$Na & 3.175 & 0.481 & 242.5\\
$^{37}$Na & 3.135 & -0.345 & 238.4 & $^{37}$Na & 3.190 & 0.512 & 243.1\\
$^{38}$Na & 3.156 & -0.359 & 240.0 & $^{38}$Na & 3.199 & 0.491 & 243.4\\
$^{39}$Na & 3.180 & -0.375 & 241.8 & $^{39}$Na & 3.209 & 0.472 & 244.1\\
$^{40}$Na & 3.184 & -0.358 & 241.3 & $^{40}$Na & 3.228 & 0.477 & 243.4\\
$^{24}$Mg & 3.043 & 0.487 & 194.3 & $^{24}$Mg & 3.001 & -0.256 & 186.8\\
$^{25}$Mg & 3.009 & 0.376 & 202.9 & $^{25}$Mg & 2.993 & -0.261 & 199.1\\
$^{26}$Mg & 2.978 & 0.273 & 212.5 & $^{26}$Mg & 2.990 & -0.268 & 211.6\\
$^{27}$Mg & 3.015 & 0.310 & 220.2 & $^{27}$Mg & 2.988 & -0.204 & 218.2\\
$^{28}$Mg & 3.048 & 0.345 & 228.7 & $^{28}$Mg & 2.992 & -0.154 & 225.6\\
$^{29}$Mg & 3.055 & 0.289 & 234.3 & $^{29}$Mg & 3.027 & -0.186 & 232.0\\
$^{30}$Mg & 3.062 & 0.241 & 240.5 & $^{30}$Mg & 3.059 & -0.207 & 239.0\\
$^{30}$Mg & 3.131 & 0.599 & 237.7 &  &  &  & \\
$^{31}$Mg & 3.075 & 0.179 & 245.1 & $^{31}$Mg & 3.068 & -0.128 & 244.0\\
$^{32}$Mg & 3.090 & 0.119 & 250.5 & $^{32}$Mg & 3.085 & -0.067 & 249.9\\
$^{32}$Mg & 3.131 & 0.471 & 248.8 &  &  &  & \\
$^{33}$Mg & 3.117 & 0.233 & 253.1 & $^{33}$Mg & 3.102 & -0.126 & 251.5\\
$^{34}$Mg & 3.150 & 0.343 & 257.3 & $^{34}$Mg & 3.124 & -0.181 & 253.9\\
$^{34}$Mg & 3.184 & 0.588 & 254.1 &  &  &  & \\
$^{35}$Mg & 3.173 & 0.388 & 260.5 & $^{35}$Mg & 3.141 & -0.196 & 255.3\\
$^{36}$Mg & 3.198 & 0.432 & 263.9 & $^{36}$Mg & 3.160 & -0.213 & 257.0\\
$^{37}$Mg & 3.212 & 0.462 & 264.9 & $^{37}$Mg & 3.179 & -0.258 & 258.8\\
$^{38}$Mg & 3.227 & 0.492 & 266.3 & $^{38}$Mg & 3.198 & -0.300 & 261.1\\
$^{39}$Mg & 3.237 & 0.473 & 267.8 & $^{39}$Mg & 3.216 & -0.338 & 263.4\\
$^{40}$Mg & 3.247 & 0.456 & 269.7 & $^{40}$Mg & 3.234 & -0.374 & 266.4\\
\hline
\hline
\end{tabular}
\end{table}

\begin{table}
\caption{\label{tab:table4}{\it Same as Table III, but for
Al, Si and S.  }}
\begin{tabular}{|c|c|c|c|c|c|c|c|c|c|}
\hline
\hline
nucleus &$r_{ch}$&$\beta_2$&BE(MeV)&nucleus &$r_{ch}$&$\beta_2$&BE(MeV)&\\
\hline
$^{24}$Al & 3.097 & 0.388 & 182.3 & $^{24}$Al & 3.077 & -0.258 & 179.4\\
$^{25}$Al & 3.072 & 0.381 & 197.7 & $^{25}$Al & 3.060 & -0.266 & 193.9\\
$^{26}$Al & 3.191 & 0.550 & 206.6 & $^{26}$Al & 3.052 & -0.275 & 207.8\\
$^{27}$Al & 3.215 & 0.572 & 217.0 & $^{27}$Al & 3.053 & -0.292 & 221.9\\
$^{28}$Al & 3.178 & 0.471 & 226.7 & $^{28}$Al & 3.037 & -0.208 & 238.6\\
$^{29}$Al & 3.061 & 0.251 & 239.3 & $^{29}$Al & 3.033 & -0.141 & 245.6\\
$^{30}$Al & 3.073 & 0.207 & 246.2 & $^{30}$Al & 3.070 & -0.184 & 253.8\\
$^{31}$Al & 3.085 & 0.170 & 253.6 & $^{31}$Al & 3.101 & -0.205 & 259.8\\
$^{32}$Al & 3.101 & 0.113 & 260.0 & $^{32}$Al & 3.103 & -0.111 & 261.2\\
$^{33}$Al & 3.118 & 0.057 & 267.2 & $^{33}$Al & 3.165 & -0.333 & 269.4\\
$^{34}$Al & 3.139 & 0.159 & 269.9 & $^{34}$Al & 3.134 & -0.108 & 275.1\\
$^{35}$Al & 3.167 & 0.268 & 274.1 & $^{35}$Al & 3.157 & -0.172 & 272.8\\
$^{36}$Al & 3.187 & 0.313 & 277.6 & $^{36}$Al & 3.173 & -0.189 & 277.7\\
$^{37}$Al & 3.208 & 0.355 & 281.5 & $^{37}$Al & 3.191 & -0.208 & 280.3\\
$^{38}$Al & 3.275 & 0.418 & 282.7 & $^{38}$Al & 3.214 & -0.254 & 283.5\\
$^{39}$Al & 3.285 & 0.406 & 285.1 & $^{39}$Al & 3.236 & -0.299 & 286.7\\
$^{40}$Al & 3.304 & 0.441 & 287.6 & $^{40}$Al & 3.257 & -0.336 & 290.4\\
$^{41}$Al & 3.325 & 0.474 & 290.5 & $^{41}$Al & 3.278 & -0.370 & 290.6\\
$^{42}$Al & 3.348 & 0.483 & 290.7 & $^{42}$Al & 3.281 & -0.355 & 291.2\\
$^{43}$Al & 3.371 & 0.491 & 291.3 & $^{43}$Al & 3.282 & -0.338 & 292.2\\
$^{44}$Al & 3.375 & 0.456 & 291.0 & $^{44}$Al & 3.274 & -0.288 & 293.6\\
$^{45}$Al & 3.378 & 0.420 & 291.0 & $^{45}$Al & 3.271 & -0.263 & 293.5\\
$^{46}$Al & 3.359 & 0.341 & 294.5 & $^{46}$Al & 3.346 & -0.296 & 294.0\\
$^{46}$Al & 3.246 & 0.125 & 293.6 & $^{46}$Al & 3.432 &  0.660 & 290.8\\
$^{47}$Al & 3.246 & 0.090 & 294.8 & $^{47}$Al & 3.335 & -0.319 & 293.6\\
$^{47}$Al & 3.447 & 0.653 & 290.4 &           &       &        &      \\
$^{48}$Al & 3.276 & 0.117 & 293.8 & $^{48}$Al & 3.319 & -0.252 & 294.0\\
$^{24}$Si & 3.162 & 0.294 & 169.1 & $^{24}$Si & 3.170 & -0.278 & 169.3\\
$^{25}$Si & 3.127 & 0.286 & 185.1 & $^{25}$Si & 3.139 & -0.274 & 184.8\\
$^{26}$Si & 3.099 & 0.282 & 201.8 & $^{26}$Si & 3.118 & -0.280 & 200.9\\
$^{27}$Si & 3.054 & 0.168 & 215.8 & $^{27}$Si & 3.114 & -0.299 & 216.4\\
$^{28}$Si & 3.017 & 0.001 & 231.4 & $^{28}$Si & 3.122 & -0.331 & 232.1\\
$^{29}$Si & 3.035 & 0.001 & 240.7 & $^{29}$Si & 3.093 & -0.237 & 240.7\\
$^{30}$Si & 3.070 & 0.148 & 250.6 & $^{30}$Si & 3.054 & -0.060 & 250.4\\
$^{31}$Si & 3.089 & 0.120 & 258.7 & $^{31}$Si & 3.108 & -0.180 & 259.1\\
$^{32}$Si & 3.109 & 0.104 & 267.2 & $^{32}$Si & 3.137 & -0.201 & 268.5\\
$^{33}$Si & 3.126 & 0.050 & 275.4 & $^{33}$Si & 3.131 & -0.084 & 275.6\\
$^{34}$Si & 3.148 & 0.000 & 284.4 & $^{34}$Si & 3.204 & -0.336 & 278.5\\
$^{35}$Si & 3.160 & 0.085 & 287.3 & $^{35}$Si & 3.161 & -0.083 & 287.4\\
$^{36}$Si & 3.184 & 0.193 & 291.4 & $^{36}$Si & 3.186 & -0.162 & 291.5\\
$^{37}$Si & 3.200 & 0.238 & 295.4 & $^{37}$Si & 3.201 & -0.181 & 294.8\\
$^{38}$Si & 3.218 & 0.281 & 299.8 & $^{38}$Si & 3.219 & -0.204 & 298.8\\
$^{39}$Si & 3.224 & 0.263 & 302.4 & $^{39}$Si & 3.245 & -0.254 & 301.9\\
$^{40}$Si & 3.232 & 0.244 & 305.4 & $^{40}$Si & 3.272 & -0.301 & 306.0\\
$^{41}$Si & 3.230 & 0.167 & 307.1 & $^{41}$Si & 3.295 & -0.336 & 310.1\\
$^{42}$Si & 3.228 & 0.013 & 309.8 & $^{42}$Si & 3.318 & -0.369 & 314.6\\
$^{43}$Si & 3.240 & 0.123 & 311.8 & $^{43}$Si & 3.320 & -0.356 & 315.2\\
$^{44}$Si & 3.252 & 0.172 & 314.3 & $^{44}$Si & 3.322 & -0.342 & 316.2\\
$^{45}$Si & 3.252 & 0.117 & 315.8 & $^{45}$Si & 3.316 & -0.308 & 317.5\\
$^{46}$Si & 3.253 & 0.053 & 317.9 & $^{46}$Si & 3.303 & -0.262 & 319.3\\
$^{47}$Si & 3.258 & 0.005 & 319.7 & $^{47}$Si & 3.345 & -0.298 & 319.8\\
$^{48}$Si & 3.263 & 0.001 & 321.8 & $^{48}$Si & 3.381 & -0.321 & 320.8\\
$^{49}$Si & 3.290 & 0.045 & 321.1 & $^{49}$Si & 3.366 & -0.251 & 320.7\\
$^{50}$Si & 3.319 & 0.074 & 321.1 & $^{50}$Si & 3.341 & -0.159 & 321.5\\
$^{51}$Si & 3.345 & 0.078 & 321.1 & $^{51}$Si & 3.358 & -0.135 & 321.2\\
$^{52}$Si & 3.371 & 0.082 & 321.4 & $^{52}$Si & 3.377 & -0.112 & 321.2\\
$^{53}$Si & 3.391 & 0.042 & 321.6 & $^{53}$Si & 3.391 & -0.052 & 321.3 \\
$^{54}$Si & 3.415 & 0.000 & 322.3 & $^{54}$Si & 3.415 & -0.010 & 322.0\\
\hline
\hline
\end{tabular}
\end{table}

\begin{table}
\caption{\label{tab:table5}{\it Same as Table III, but for S.
}}
\begin{tabular}{|c|c|c|c|c|c|c|c|c|c|}
\hline
\hline
nucleus &$r_{ch}$&$\beta_2$&BE(MeV)&nucleus &$r_{ch}$&$\beta_2$&BE(MeV)\\
\hline
$^{33}$S & 3.241 & 0.197 & 275.5 & $^{33}$S & 3.233 & -0.116 & 275.1\\
$^{34}$S & 3.248 & 0.140 & 285.8 & $^{34}$S & 3.257 & -0.168 & 286.5\\
$^{35}$S & 3.260 & 0.077 & 295.6 & $^{35}$S & 3.260 & -0.078 & 295.7\\
$^{36}$S & 3.273 & 0.002 & 306.2 & $^{36}$S & 3.309 & -0.308 & 299.7\\
$^{37}$S & 3.285 & 0.152 & 311.6 & $^{37}$S & 3.287 & -0.116 & 310.1\\
$^{38}$S & 3.300 & 0.228 & 318.6 & $^{38}$S & 3.300 & -0.164 & 316.9\\
$^{39}$S & 3.312 & 0.264 & 325.3 & $^{39}$S & 3.307 & -0.173 & 322.6\\
$^{40}$S & 3.325 & 0.299 & 332.4 & $^{40}$S & 3.316 & -0.181 & 328.5\\
$^{41}$S & 3.331 & 0.287 & 337.7 & $^{41}$S & 3.324 & -0.189 & 333.6\\
$^{42}$S & 3.338 & 0.277 & 343.2 & $^{42}$S & 3.335 & -0.207 & 339.2 \\
$^{43}$S & 3.359 & 0.318 & 347.2 & $^{43}$S & 3.348 & -0.229 & 344.1\\
$^{44}$S & 3.381 & 0.367 & 351.0 & $^{44}$S & 3.366 & -0.263 & 349.5\\
$^{45}$S & 3.375 & 0.312 & 353.4 & $^{45}$S & 3.367 & -0.240 & 351.8\\
$^{46}$S & 3.371 & 0.258 & 356.6 & $^{46}$S & 3.375 & -0.237 & 355.1\\
$^{47}$S & 3.385 & 0.257 & 358.5 & $^{47}$S & 3.380 & -0.230 & 358.0\\
$^{48}$S & 3.400 & 0.259 & 360.8 & $^{48}$S & 3.389 & -0.230 & 360.8\\
$^{49}$S & 3.403 & 0.227 & 362.9 & $^{49}$S & 3.420 & -0.257 & 362.9\\
$^{50}$S & 3.403 & 0.189 & 365.5 & $^{50}$S & 3.451 & -0.277 & 365.0\\
$^{51}$S & 3.427 & 0.188 & 366.4 & $^{51}$S & 3.451 & -0.231 & 365.9\\
$^{52}$S & 3.451 & 0.183 & 367.6 & $^{52}$S & 3.447 & -0.178 & 367.6\\
$^{53}$S & 3.463 & 0.158 & 369.1 & $^{53}$S & 3.466 & -0.172 & 368.4\\
$^{54}$S & 3.477 & 0.139 & 371.0 & $^{54}$S & 3.486 & -0.142 & 369.8\\
$^{55}$S & 3.494 & 0.105 & 371.4 & $^{55}$S & 3.497 & -0.090 & 370.5\\
\hline
\hline
\end{tabular}
\end{table}

\begin{table}
\caption{\label{tab:table6}{\it 
The calculated value of charge radii ($r_{ch}$),
quadrupole deformation parameter $\beta_2$ and binding energy (BE) for
Ne, Mg, Si and S even-even nuclei in SHF (SkI4) formalism. The maximum 
binding energy is the
ground state solution and all other values are the intrinsic excited state
solution. The radius $r_{ch}$ is in fm and the binding energy is in MeV.
}}
\begin{tabular}{|c|c|c|c|c|c|c|}
\hline
\hline
nucleus & $r_{ch}$& $\beta_2$ & BE(MeV)& $r_{ch}$ & $\beta_2$ &BE(MeV) \\
\hline
$^{20}$Ne & 3.029 & 0.5481 & 156.817 & 2.950 &-0.1356 & 154.474 \\
$^{22}$Ne & 3.005 & 0.5223 & 175.758 & 2.943 & -0.1989 & 172.758 \\
$^{24}$Ne & 2.952 & 0.2546 & 188.354 & 2.951 & -0.2541 & 188.538 \\
$^{26}$Ne & 2.953 & 0.1233 & 199.380 & 2.944 & 0.0060 & 199.389 \\
$^{28}$Ne & 3.013 & 0.1623 & 206.524 & 3.010 & -0.1334 & 206.433 \\
$^{30}$Ne & 3.054 & 0.0030 & 213.721 &  &  &   \\
$^{32}$Ne & 3.103 & 0.3808 & 213.118 & 3.118 & 0.3716 & 213.215 \\
$^{34}$Ne & 3.179 & 0.4880 & 213.483 & 3.108 & -0.1462 & 209.695 \\
$^{36}$Ne & 3.203 & 0.6015 & 212.230 & 3.147 & -0.2789 & 208.770 \\
$^{24}$Mg & 3.128 & 0.5248 & 195.174 & 3.077 & -0.252 & 189.946  \\
$^{26}$Mg & 3.090 & 0.3623 & 212.885 & 3.079 & -0.2988 & 213.153 \\
$^{28}$Mg & 3.111 & 0.3419 & 228.997 & 3.056 & -0.1076 & 227.899 \\
$^{30}$Mg & 3.119 & 0.2022 & 240.328 & 3.117 & -0.1835 & 240.514 \\
$^{32}$Mg & 3.145 & 0.0000 & 252.033 & 3.145 & 0.0000 & 252.033  \\
$^{34}$Mg & 3.209 & 0.3263 & 255.067 & 3.175 & -0.1196 & 253.455 \\
$^{35}$Mg & 3.295 & 0.4884 & 201.512 & 3.252 & -0.289 & 257.634  \\
$^{36}$Mg & 3.265 & 0.4413 & 259.899 & 3.213 & -0.2124 & 255.368 \\
$^{40}$Mg & 3.321 & 0.4741 & 262.796 & 3.299 & -0.3538 & 260.200 \\
$^{28}$Si & 3.117 & 0.009 & 231.037 & 3.194 &  -0.3494 & 233.590  \\
$^{30}$Si & 3.145 & 0.1477 & 252.146 & 3.168 & -0.2102 & 252.625 \\
$^{32}$Si & 3.179 & 0.007 & 269.479 & 3.199 & -0.1990 & 270.483 \\
$^{34}$Si & 3.216 & 0.000 & 286.332 &  & & \\
$^{36}$Si & 3.146 & 0.1549 & 292.418 & 3.241 & -0.009 & 292.425 \\
$^{38}$Si & 3.291 & 0.3051 & 298.173 & 3.279 & -0.1978& 298.173 \\
$^{40}$Si & 3.325 & 0.2990 & 230.450 & 3.309 & -0.2817 & 303.969 \\
$^{42}$Si & 3.349 & 0.3592 & 307.399 & 3.334 & -0.3508 & 310.023 \\
$^{44}$Si & 3.334 & 0.2119 & 309.712 & 3.377 & -0.3031 & 311.601  \\
$^{46}$Si & 3.337 & 0.009 & 312.451 & 3.372 & -0.2405 & 313.508 \\
$^{48}$Si & 3.348 & 0.002 & 315.425 & 3.438 & -0.2893 & 313.995 \\
$^{30}$S & 3.262 & 0.1491 & 241.434 & 3.178 & -0.1856 & 241.385 \\
$^{32}$S & 3.271 & 0.200  & 208.173 & 3.256 & -0.1700 & 267.975 \\
$^{34}$S & 3.288 & 0.1212 & 288.804 & 3.295 & -0.1566 & 289.304 \\
$^{36}$S &  &  & & 3.314 & -0.003 & 309.619 \\
$^{38}$S & 3.341 & 0.2144 & 320.168 & 3.331 & -0.1289 & 318.951 \\
$^{40}$S & 3.374 & 0.3042 & 332.097 & 3.350 & -0.1529 & 327.809 \\
$^{42}$S & 3.392 & 0.2898 & 341.033 & 3.379 & -0.2195 & 337.031 \\
$^{44}$S & 3.436 & 0.3677 & 348.266 & 3.410 & -0.2714 & 346.445  \\
$^{46}$S & 3.423 & 0.2517 & 352.486 & 3.413 & -0.2090 & 351.578 \\
$^{48}$S & 3.450 & 0.2379 & 356.188 & 3.430 & -0.2032 & 356.589  \\
$^{50}$S & 3.441 & 0.1229 & 360.815 & 3.498 & -0.2673 & 359.011 \\
$^{52}$S & 3.482 & 0.1099 & 362.347 & 3.487 & -0.1356 & 362.531 \\
$^{54}$S & 3.528 & 0.003  & 364.650 & 3.524 & -0.1023 & 363.926 \\
$^{56}$S & 3.558 & 0.1105 & 366.031 & 3.556 & -0.0100 & 366.033\\
\hline
\hline
\end{tabular}
\end{table}

The drip-lines are very important after discovery of 
the two isotopes $^{40}$Mg and $^{42}$Al \cite{baumann07} that here once
predicted 
to be beyond the drip-line \cite{moller95,samyn04}. This suggests that the 
drip-line is somewhere in the heavier side
of the mass prediction which are beyond the scope of the present mass models 
\cite{moller95,samyn04}. In this calculations the newly discovered
nuclei $^{40}$Mg and $^{42}$Al are well within the prediction both 
in the SHF and RMF formalisms.
Again a further comparison of the drip-line with RMF and SHF prediction,
we find the drip-line predictions in both calculations are well comparable,
except for a few exceptions in Na and Si as shown in Table II.

\subsection{Neutron configuration}
Analysing the neutron configuration for these exotic nuclei, we notice 
that, for lighter isotopes
of Ne, Na, Mg, Al, Si and S the oscillator shell $N_{osc}=3$ is empty. 
However, the $N_{osc}=3$  shell gets occupied gradually with increase of 
neutron number. In case of Na, $N_{osc}=3$ starts filling up at $^{33}$Na with quadrupole
moment deformation parameter $\beta_2=0.356$ and $-0.179$ with occupied orbits
$[330]1/2^{-}$ and $[303]7/2^{-}$, respectively. The filling of $N_{osc}=3$  
goes on
increasing for Na with neutron number and it is $[330]1/2^{-}$, 
$[310]1/2^{-}$, $[321]3/2^{-}$ and $[312]5/2^{-}$ at $\beta_2=0.472$ for 
$^{39}$Na.
Again for the oblate solution the occupation is  $[301]1/2^{-}$,
$[301]3/2^{-}$, $[303]5/2^{-}$ and $[303]7/2^{-}$ for $\beta_2=-0.375$
for $^{39}$Na.
In the case of Mg isotopes, even for $^{30,32}$Mg, the $N_{osc}=3$ shell have
some occupation for the low-lying excited states near the Fermi surface for $^{30}Mg$ (at
$\beta$= 0.599 with Be = 237.721 MeV the $N_{osc}$=3 orbit is $[330]7/2^{-}$ and for $^{32}Mg$: 
$\beta_2=[330]1/2^{-}$, BE=248.804 MeV at $\beta_2=0.471$). With the increase
of neutron number in Mg and Si isotopic chain, the oscillator
shell with $N_{osc}=3$ gets occupied more and more.  For most of the 
Si isotopes, the oblate solutions are the dominating ones than the low-lying 
prolate superdeformed states, i.e. mass of the oblate solutions are the ground state 
solutions and the prolate and some superdefomed are the excited configurations. Again, in 
S-isotopes, the prolate are the ground state and the oblate are the extreme excited states.  
Note that in many cases, there exist low laying superdeformed states and some of them are 
listed in the Tables.

\subsection{Quadrupole deformation}
The ground and low-lying excited state deformation systematics for some 
of the representative nuclei for Ne, Na, Mg, Al, Si and S are analysed.
In Fig. 1, the ground state quadrupole deformation parameter $\beta_2$ 
is shown as a function of mass number for Ne, Na, Mg, Al, Si and S. 
The $\beta_2$ value goes on increasing with mass number for Ne, Na and Mg
isotopes near the drip-line. 
The calculated quadrupole deformation parameter 
$\beta_2$ for $^{34}$Mg is 0.59 which compares well with the recent 
experimental measurement of Iwasaki et al \cite{iwasaki01} 
($\beta_2=0.58\pm 6$). Note that this superdeformed states in 3.2 MeV  
above than the ground band. Again, the magnitude of $\beta_2$ for 
the drip-line nuclei reduces with neutron number N and again increases. 
A region of maximum deformation is found for almost all the nuclei
as shown in the figure. It so happens in cases like, Ne, Na, Mg and
Al that the isotopes are maximum deformed which may be comparabled to
superdeformed near the drip-line. For Si isotopes, 
in general, we find oblate solution in the ground configurations. In many
of the cases, the low-lying superdeformed configuration are clearly
visible and some of them are available in the Tables.  
\begin{figure}
\includegraphics[scale=0.3]{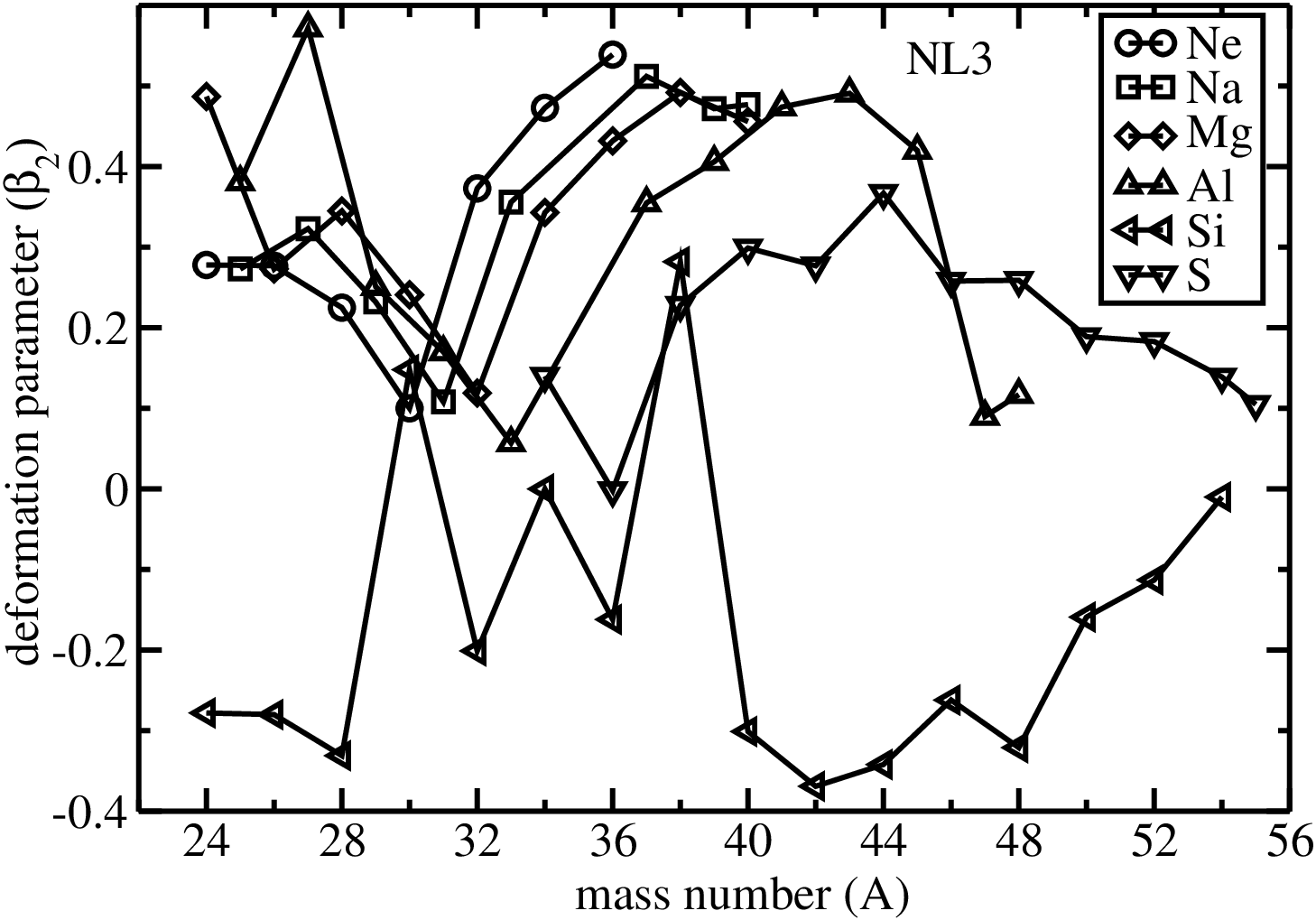}
\caption{\label{fig:epsart}{\it The ground state quadrupole deformation
parameter $\beta_2$ versus mass number A for Ne, Na, Mg, Al, Si and S isotopes
near the drip-line with NL3 parameter set.
}}
\end{figure}

\subsection{Shape coexistence}
One of the most interesting phenomena in nuclear structure physics is the
shape coexistence \cite{maharana92,sarazin00,egido04}. In many of the cases 
for the nuclei  considered here near the drip-line isotopes, the ground state 
configuration accompanies a low-lying excited state. In few cases, it so 
happens that these two solutions
are almost degenerate. That means we predict almost similar binding energy
for two different configurations. For example, in the RMF calculation,
the ground state binding energy
of $^{24}$Ne is 189.093 MeV with $\beta_2=-0.259$ and the binding energy of 
the excited low-lying configuration at $\beta_2=0.278$ is 188.914 MeV. 
The difference
in BE of these two solutions is only 0.179 MeV. Similarly the
solution of prolate-oblate binding energy difference in SkI4 is 0.186 MeV
for $^{30}$Mg with $\beta_2=-0.183$ and 0.202. This phenomenon is
clearly available in most of the isotopes near the drip-line.  
\begin{figure}
\includegraphics[scale=0.3]{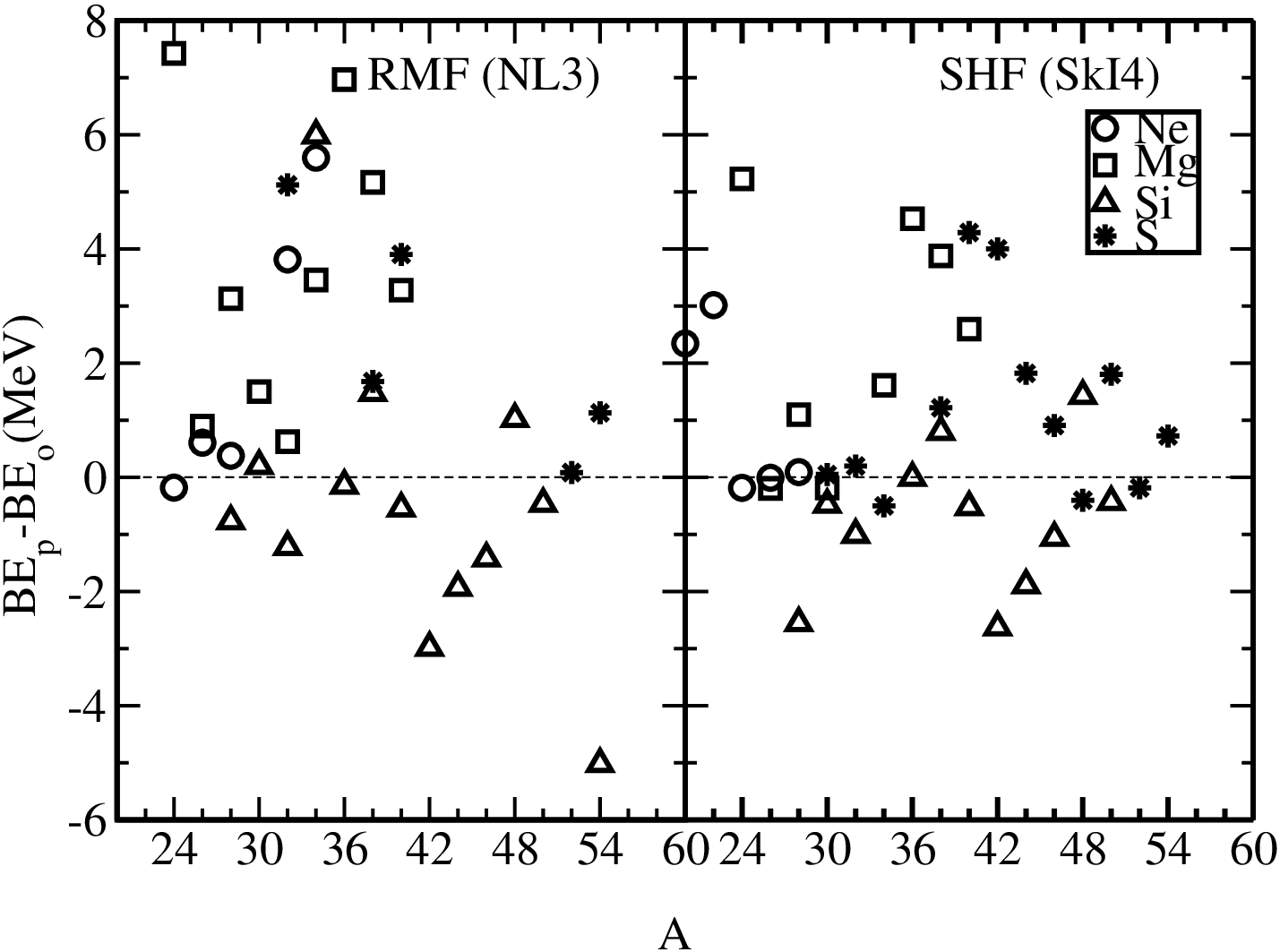}
\caption{\label{fig:epsart}{\it The difference in binding energy 
between the prolate-oblate solutions is shown for even-even Ne, Mg, Si and S
isotopes near the neutron drip-line with NL3 and SkI4 parameter sets.
}}
\end{figure}
To show it in a more quantitative way, we have plotted the prolate-oblate
binding energy difference in Figure 2. The left hand side of the figure
is for relativistic and the right side is the nonrelativistic SkI4 results.
From the figure, it is clear that an island of shape coexistence
isotopes are available for Mg and Si isotopes. These shape coexistence
solutions are predicted taking into account the intrinsic binding energy.
However the actual quantitative energy difference of ground and excited 
configuration can be given by performing
the angular momentum projection, which is be an interesting problem for
future.   

\begin{figure}
\includegraphics[scale=0.3]{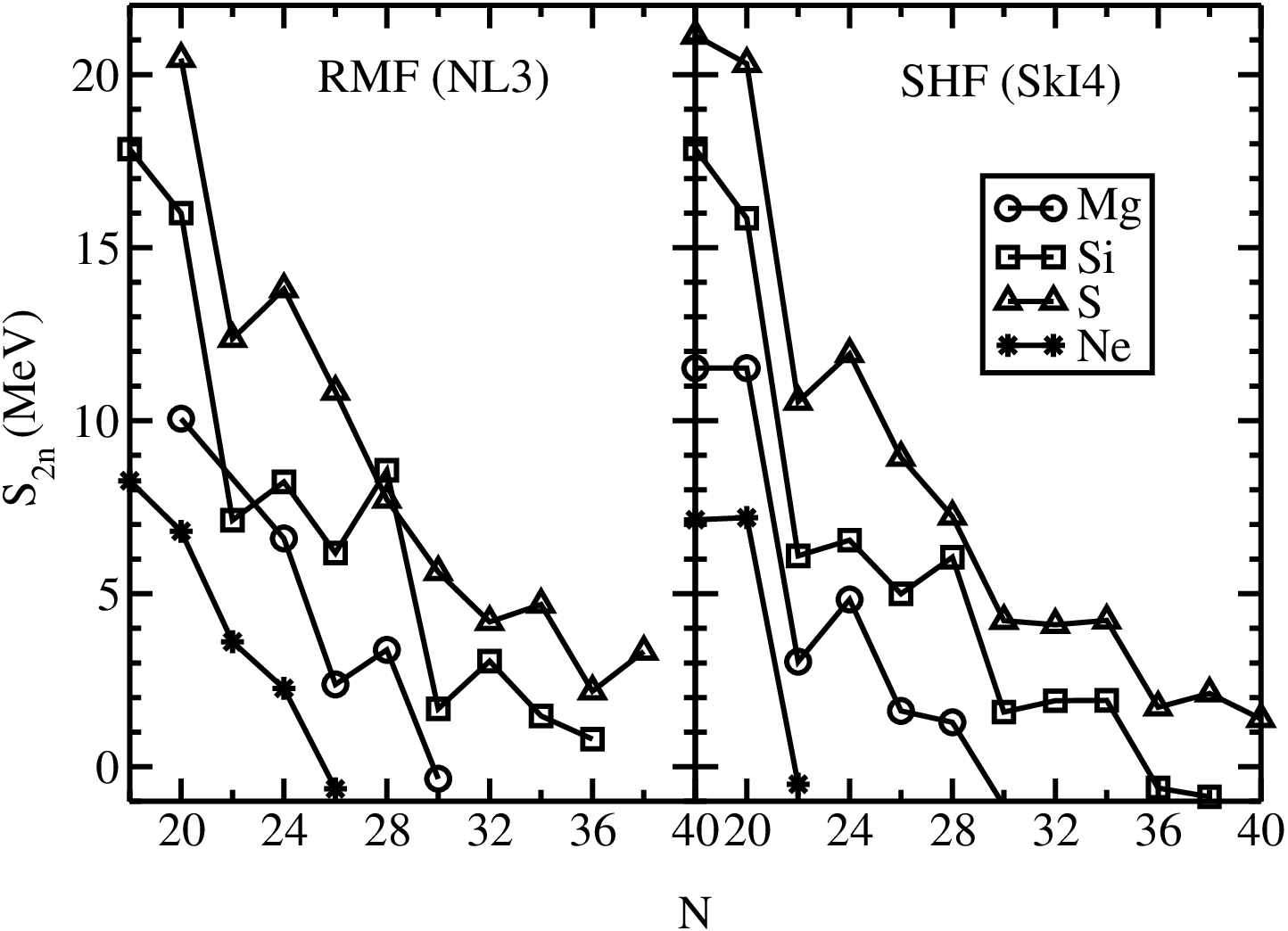}
\caption{\label{fig:epsart}{\it The two-neutron separation $S_{2n}$ energy
versus neutron number N for neutron-rich Ne, Mg, Si and S isotopes. 
}}
\end{figure}
\subsection{Two neutron separation energy and new magic number}

The appearance of new and the disappearance of known
magic number near the neutron drip-line is a well discussed topic 
currently in nuclear structure physics \cite{ozawa00,jha03}. Some of the
calculations in recent past predicted the disappearance of the known magic
number N=28 for the drip-line isotopes of Mg and S \cite{ren01,werner94}. 
However, magic number 20 retains its magic properties even for the drip-line
region.
In one of our earlier publications, \cite{patra} we analysed the spherical shell gap
at N=28 in $^{44}$S and its neighboring $^{40}$Mg and $^{42}$Si using NL-SH
\cite{sharma93} and TM2 parameter sets \cite{toki94}. The spherical shell 
gap at N=28 in $^{44}$S was found to be intact for the TM2 and is broken for 
NL-SH parametrization.  
Here, we plot the two-neutron separation energy $S_{2n}$ of Ne, 
Mg, Si and S for the even-even nuclei near the drip-line (fig 3). The
known magic number N=28 is noticed to be absent in $^{44}$S. On the
other hand the appearance of steep 2n-separation energy at N=34 both 
in RMF and SHF calculation
looks quite prominent, and this is just two units ahead than the experimental
shell closure N=32 \cite{kanungo02}.

\subsection{ Superdeformation and Low $\Omega$ parity doublets}

The deformation-driving $m=1/2-$orbits come down in energy
in superdeformed solutions from the shell above, in contrast to
the normal deformed solutions. The occurrence of approximate
$1/2^+$ $1/2^-$ parity dobulets (degeneracy of $|m|^{\pi}$=
$1/2^+$ $1/2^-$ states) for the superdeformed solutions are clearly
seen in Figs. 4 and 5 where excited superdeformed configurations
for $^{32}$Mg and $^{34}$Mg and for $^{46}$Al and $^{47}$Al are given.
For each nucleus we have compared the normal deformed
$(\beta_{2} \sim 0.1-0.3)$ and the superdeformed configurations and 
analysed the deformed orbits.
\begin{figure}
\includegraphics[scale=0.4]{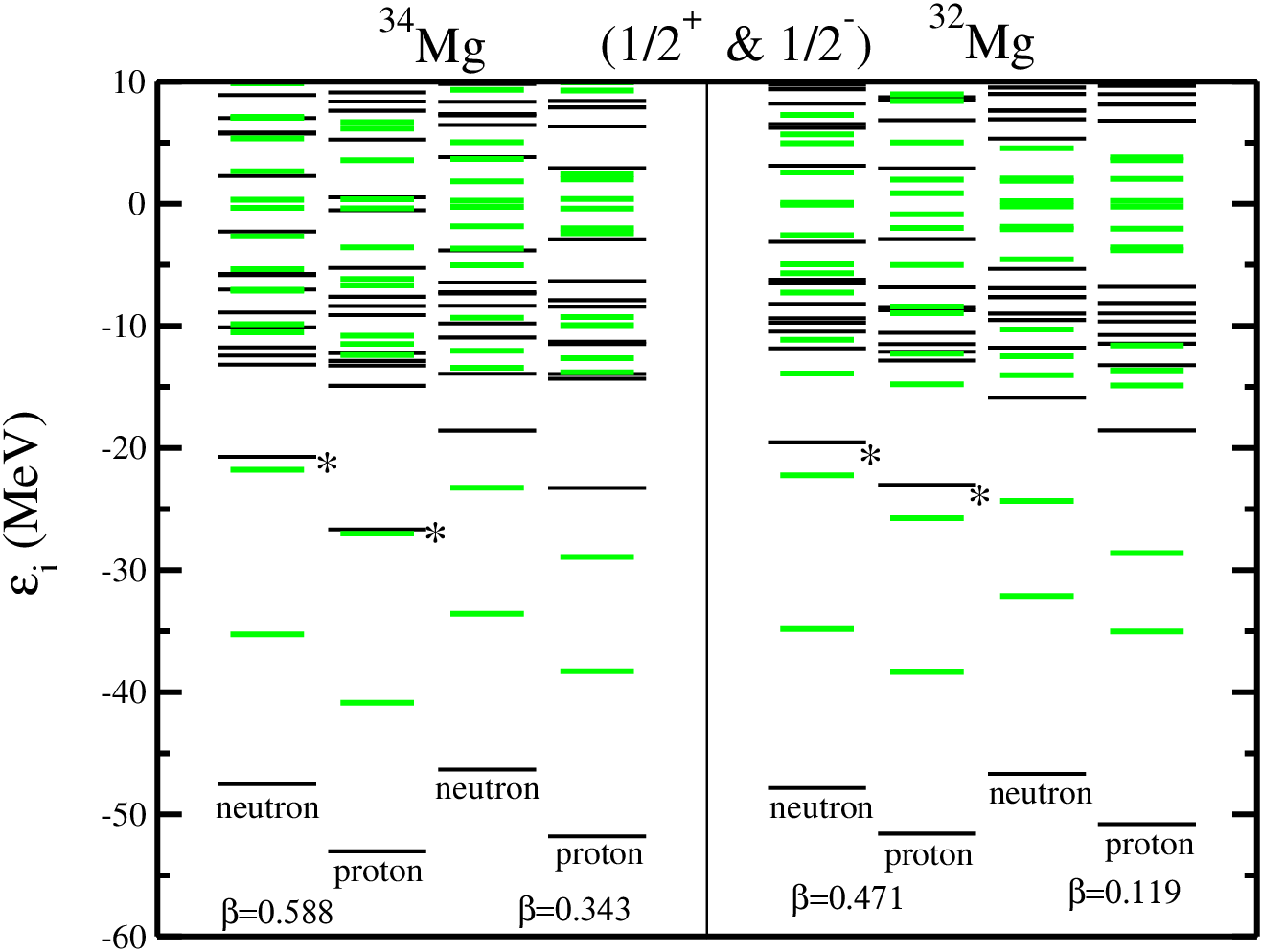}
\caption{\label{fig:epsart}{\it The $1/2^+ and 1/2^-$ intrinsic 
single-particle states for
the normal and superdeformed state for $^{32}$Mg and $^{34}$Mg. A few of the 
lowest energy parity-doublet states of the superdeformed (SD) solutions are 
shown by asterisk for the SD configuration. More such doublets are noticed 
for the SD intrinsic states. The ${\pm 1/2^-}$ states are denoted by shorter
(and green) lines and the ${\pm 1/2^+}$ states are denoted by longer (
and black).  
}}
\end{figure}
\begin{figure}
\includegraphics[scale=0.4]{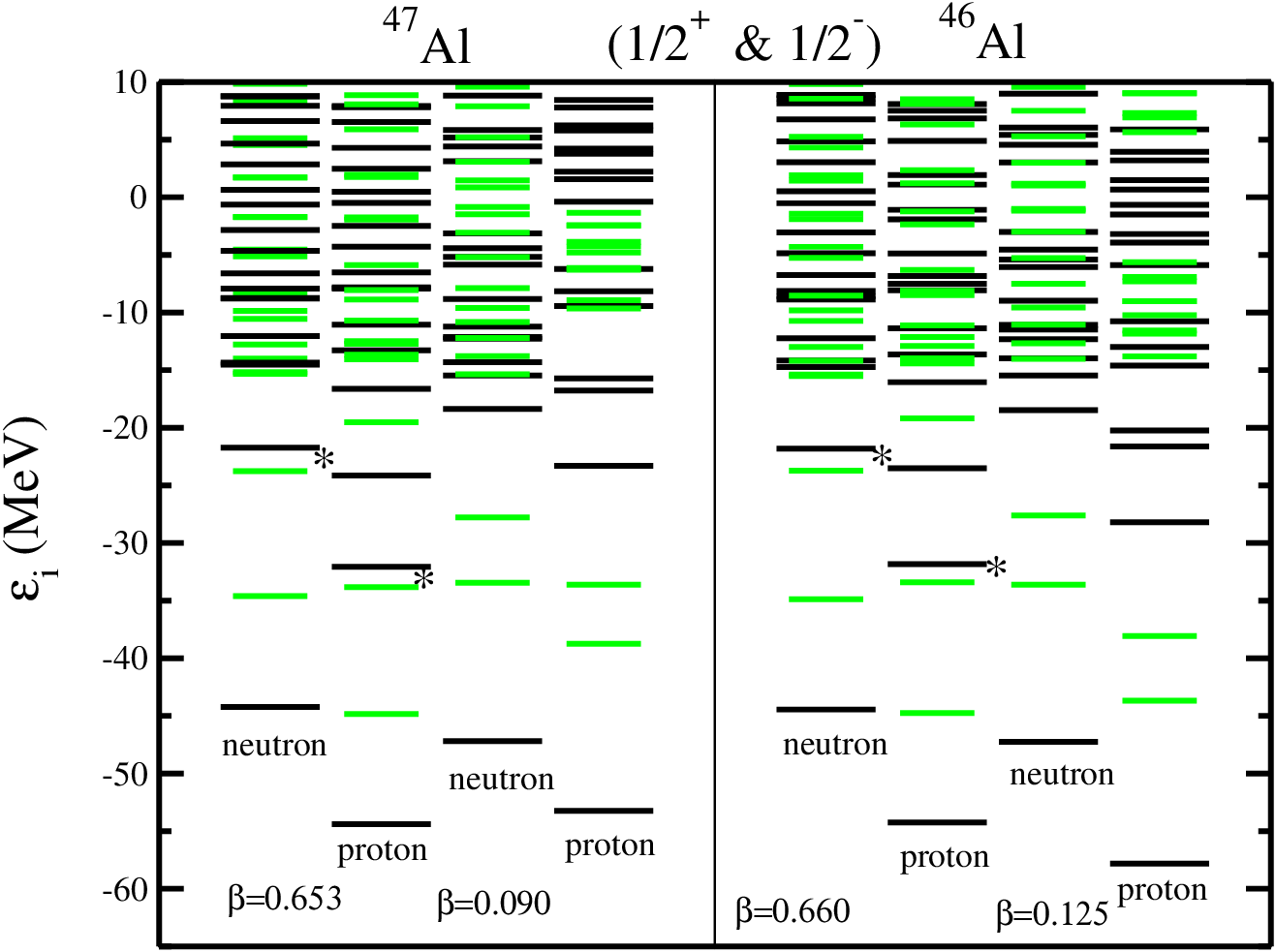}
\caption{\label{fig:epsart}{\it Same as Fig. 4 for $^{46}$Al 
and $^{47}$Al. 
}}
\end{figure}

The $1/2^+$ and $1/2^-$ states for the single particle levels are
shown in Fig. 4 (for $^{32}$Mg and $^{34}$Mg). From the analysis of 
the results of this calculation,
we have found a systematic
behaviour of the low $\Omega$ (particularly $1/2^+$ and $1/2^-$) prolate
deformed orbits for the superdeformed solutions. As representive cases, 
we present here results
for ($^{34}$Mg $-$ $^{32}$Mg) and ($^{47}$Al $-$ $^{46}$Al) 
and plot the $1/2^+$ and $1/2^-$ orbits for
the superdeformed and normal deformed shapes of these nuclei. We notice
from the plot of the orbits that there is occurrence of $1/2^+$ and $1/2^-$
orbits very closeby in energy for the superdeformed (SD) shape. 
Two such $1/2^+$ and $1/2^-$ doublet structures, marked in asterisk
are shown in Figs. 4 and 5 for the SD solutions.
Such $1/2^+$ $1/2^-$ degenerate orbits  occur not only for the
well-bound orbits but also for the unbound continuum states. As example,
the doublet neutrons $[220]1/2^+$ and $[101]1/2^-$ states is 4 MeV apart 
in energy in the normal deformed prolate 
solutions tend to become degenerate in the SD solution $[220]1/2^+$  
and $[101]1/2^-$ states (prolate) belonging to two different major shells,
so close to each other in the superdeformed solution (shown in
Figs. 4 for $^{34}$Mg). More such doublets are easily identified 
(Figs. 4 and 5) for superdeformed solutions of $^{32,34}$Mg and $^{46,47}$Al.
In fact it is noticed that the $\Omega=1/2$ states of unique parity, seen
clearly well separated in the normal deformed solutions, get quite
close to each other for the SD states, suggesting degenerate parity 
doublet structure. This can lead
to parity mixing and octupole deformed shapes for the SD structures
\cite{crp07}. Parity doublets and octupole deformation for superdeformed 
solutions have been discussed for neutron-rich Ba and Zr nuclei 
\cite{crp86,crp99}.
There is much interest for the experimental study of the spectra of
neutron-rich nuclei in this mass region \cite{miller}. The highly deformed
structures for the neutron-rich $Ne-Na-Mg-Al$ nuclei are interesting
and signature of such superdeformed configurations should be looked for.

\section{Summary and Conclusion}

In summary, we calculate the ground and low-lying excited state properties,
like binding energy and quadrupole deformation $\beta_2$ using NL3 parameter
set for Ne, Na, Mg, Si and S isotopes, near the neutron drip-line region.
In general, we find large deformed solutions for the neutron-drip 
nuclei which agree well with the experimental measurement. We have 
done the calculation using the nonrelativistic Hartree-Fock formalism with 
Skyrme 
interaction SkI4. Both the relativistic and non-relativistic results were found 
comparable to each other for the considered mass region. In the
present calculations a large number of low-lying intrinsic superdeformed
excited states are observed for many of the isotopes and some of them are reported. 
The breaking of N=28 magic number and the appearance of a new magic number at 
N=34 appears in our calculations. A proper angular momentum projection
may tell us the exact lowering of binding energy and it may happen that
the superdeformed would be the ground band of some of the neutron-rich 
nuclei. Work
in this direction is worth doing because of the present interest in
the topic of the drip-line nuclei.
In this study we find that, for the SD shape, the low $\Omega$ orbits (particularly $\Omega=1/2$)
become more bound and show a parity doublet structure.
Closelying parity-doublet band structures and enhanced electromagnetic
transition rates are a clear possibility for the
superdeformed shapes.

\section{Acknowledgments}

This work has been supported in part by Council of Scientific $\&$ Industrial 
Research (No. 03(1060)06/EMR-II) as well as projects No. SR/S2/HEP-16/2005 and
SR/S2/HEP-037/2008,
Department of Science and Technology, Govt. of India.

\end{document}